\documentclass[acmsmall, authorversion, nonacm]{acmart} 

\AtBeginDocument{%
  \providecommand\BibTeX{{%
    \normalfont B\kern-0.5em{\scshape i\kern-0.25em b}\kern-0.8em\TeX}}}

\setcopyright{acmcopyright}
\copyrightyear{2020}
\acmYear{2020}
\acmDOI{10.1145/1122445.1122456}

\usepackage{times}
\usepackage{helvet}
\usepackage{courier}
\usepackage{bigstrut}
 \usepackage{multirow}
\usepackage{tikz}
\usepackage{algorithm2e}
\usepackage{booktabs,caption}
\usepackage{lipsum}
\usepackage{graphicx}
\usepackage{array}
\usepackage{multirow}
\usepackage{caption}
\usepackage{subcaption}
\usepackage{colortbl}
\usepackage[flushleft]{threeparttable}

\begin{document}

\title{Bot-Match: Social Bot Detection with Recursive Nearest Neighbors Search}

\author{David M. Beskow}
\email{dmbeskow@gmail.com}
\orcid{0000-0003-2814-8712}
\author{Kathleen M. Carley}
\email{kathleen.carley@cs.cmu.edu}
\affiliation{%
  \institution{Carnegie Mellon University}
  \streetaddress{5000 Forbes Ave}
  \city{Pittsburgh}
  \state{PA}
  \postcode{15213}
}

\renewcommand{\shortauthors}{Beskow and Carley}

\begin{abstract}
Social bots have emerged over the last decade, initially creating a nuisance while more recently used to intimidate journalists, sway electoral events, and aggravate existing social fissures.  This social threat has spawned a bot detection algorithms race in which detection algorithms evolve in an attempt to keep up with increasingly sophisticated bot accounts.  This cat and mouse cycle has illuminated the limitations of supervised machine learning algorithms, where researchers attempt to use yesterday's data to predict tomorrow's bots.  This gap means that researchers, journalists, and analysts daily identify malicious bot accounts that are undetected by state of the art supervised bot detection algorithms.  These analysts often desire to find similar bot accounts without labeling/training a new model, where similarity can be defined by content, network position, or both.  A similarity based algorithm could complement existing supervised and unsupervised methods and fill this gap.  To this end, we present the Bot-Match methodology in which we evaluate social media embeddings that enable a semi-supervised recursive nearest neighbors search to map an emerging social cybersecurity threat given one or more seed accounts.  

\end{abstract}

\keywords{graph embedding, social bot detection, coneten based information retrieval}

\maketitle

\section{Introduction}

\noindent Today, sophisticated state and non-state actors are using information systems in general and social media in particular to change the beliefs and actions of target societies and cultures.  These (dis)information campaigns, if left unchecked, gradually degrade the target society by eroding key institutions and values while widening existing fissures.  This information ``blitzkrieg" has led to the emerging discipline of social cybersecurity in which societies attempt to protect their culture and values from external manipulation while maintaining a free market for opinions and ideas.  One of the key functions that computer science brings to the multi-disciplinary table of social cybersecurity is bot/cyborg/sybil/troll detection and characterization.  


Supervised and unsupervised machine learning models both provide important contributions to bot detection, but are not sufficient for social cybersecurity practitioners.  Supervised models trained on prevalent bot data provide an initial ``triage'' of social media streams, identifying likely areas of bot involvement and artificial manipulation of the online conversation.  However, building supervised machine learning algorithms for every bot-detection scenario quickly grows untenable.  Myriads of bot genres have evolved, including spam bots, intimidation bots, propaganda bots, social influence bots, cyborg accounts, and many others.  Each of these bot genres have unique features and are curated and deployed in various ways depending on the target audience and culture.  It is neither possible to train a single model that generalizes to every genre, nor is it convenient to label and train models for every genre and then update these models on a frequent basis to keep up with bot evolution.  Unsupervised learning, on the other hand, provides a way to find certain types of bots, such as the correlated bots found by the Debot algorithm \cite{chavoshi2016debot}.  These types of models are especially helpful in identifying labeled data for supervised models \cite{beskowOSINT2020}, but once again they are not sufficient.  Often the most sophisticated and influential dis-information bots/cyborgs can fly ``under the radar,'' undetected by either supervised or unsupervised models.  When we began to triage external manipulation of the Canadian political conversation in the runup to the Canadian 2019 national elections we found multiple influential accounts had emerged that were not being detected by production bot detection algorithms, but nonetheless were 1) divisive, 2) appeared to have foreign connections, and 3) appeared to have automated activity (i.e. were bots).  

We find that social cybersecurity practitioners in journalism, industry, academia and government, when faced with these sophisticated accounts, naturally ask the simple question ``I wonder how many other accounts are similar to this one?''  We developed \emph{Bot-Match} to fill this gap, allowing analysts to rapidly find similar accounts in a flexible manner where the analyst can determine how they want to define similarity.  This approach is complementary to supervised/unsupervised models, and is not designed to replace them.  

\begin{figure}[htb]
\centering
  \includegraphics[width=\columnwidth, trim={0cm 0cm 0cm 0cm},clip]{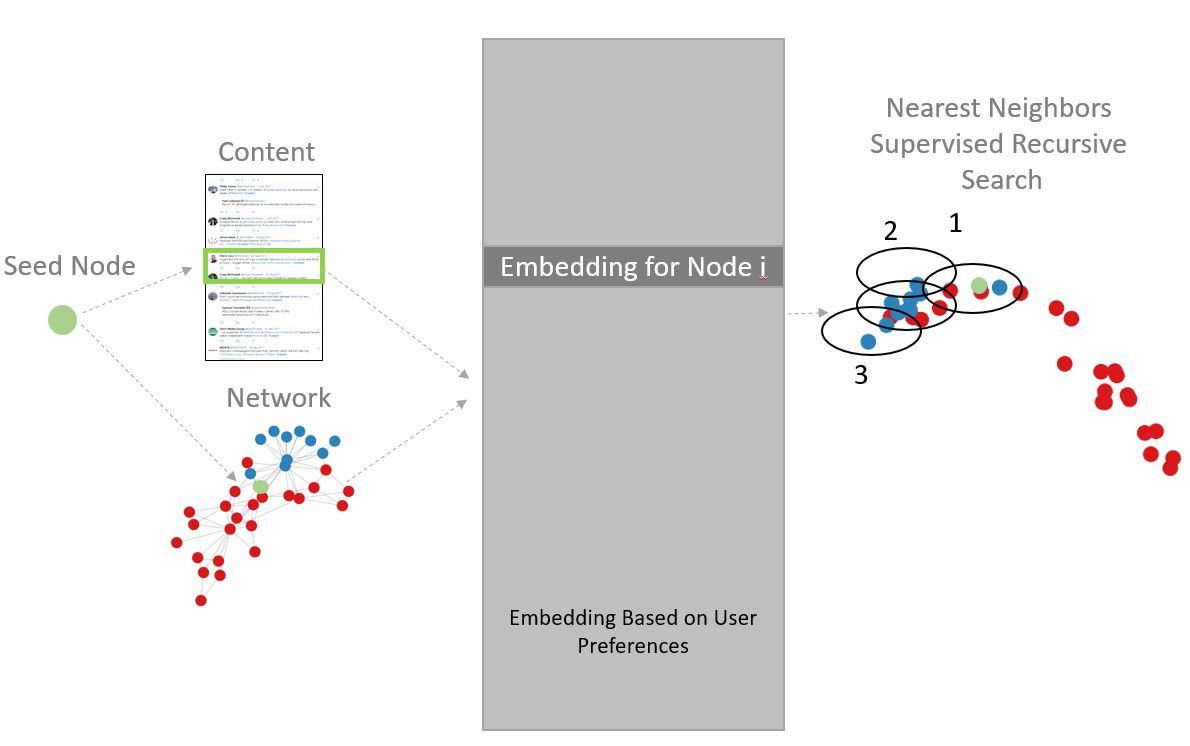}
  \caption{Framework to develop social media embeddings that enable a semi-supervised recursive nearest neighbors search to find similar accounts given one or more seed accounts}
  \label{fig:framework}
\end{figure}

Similarity could be defined as similar network connections (either similar connections or similar network role), similar content, or a combination of both.  With Bot-Match the analyst can choose to embed the conversation network, the conversation content, or both simultaneously, and then find similar accounts given a query.  In this case the query is the seed node(s), and the algorithm returns the nearest neighbors given the predefined similarity measure.  By recursively making this query, the analyst can rapidly build out a sophisticated information campaign that is undetected by other social cybersecurity tools.  This approach is illustrated in Figure \ref{fig:framework}.

Bot-match is a type of information retrieval where the query is a malicious actor and all of their features (semantic and network features).  Within information retrieval this type of query is often called Content Based Information Retrieval (CBIR). Google Reverse Image Search is an example of CBIR.  The power of Google reverse image search is that the user can upload a feature rich image rather than a general and limited semantic query.  For example, if an intelligence analyst wanted to find emerging images of terrorists, simply typing "terrorist" into the Google image search would reveal many stock photos of a stereotypical terrorist, which is not helpful.  However, if the same intelligence analyst conducted a CBIR-based reverse image search with a photo of a terrorist fighter from a specific terrorist organization, complete with distinguishing flags, background, and unique face masks, this feature rich query would provide a meaningful and relevant query result.  

Bot-match, similar to Google reverse image lookup, seeks to use the feature rich information about an account to serve as the query for information retrieval, returning a ranked list of similar accounts.  This search becomes a semi-supervised search as the user identifies additional accounts of interest, and recursively searches with these new accounts.  While recursive, in practical social cybersecurity workflows the user will seldom recurse more than a few `hops', and will not fully explore the entire stream.  

Bot-Match is designed to complement but not replace or be compared to supervised models.  Supervised models conduct a classification task, often on large streams of data.  Bot-match conducts content based information retrieval which returns a ranked list, often in a recursive manner, to explore accounts of interest and find similar accounts.  Although semi-supervised, Bot-Match also differs in purpose from other semi-supervised label/beief propagation models.  Label/belief propagation is attempting to classify a large stream with only a few labeled instances.  Some label/belief propagation methods produce a ranked list and have seen use in information retrieval.  We will analyze these below.

The primary contribution of this paper is describing content based information retrieval with a feature rich account object for social cybersecurity.  We merge ideas and models from information retrieval, graph embedding, semantic embedding, semi-supervised learning, cluster analysis, and link prediction to create a novel solution to a common hard question.   We introduce two new social cybersecurity data sets and devise a test and evaluation for candidate models.  This paper also validates this approach on social media associated with US and Canadian social media data related to election events.  While applying this in the specific context of malicious disinformation operations, other intelligence and commercial data workflows could leverage content based information retrieval with rich account features.  

This paper is organized as follows.  It begins by describing past research content based information retrieval, particularly focused in semantic and graph embedding.  We also briefly discuss relevant research from semi-supervised learning, information retrieval, and link prediction. We conduct the formal evaluation of these models on two bench mark datasets as well as two labeled data sets associated with social cybersecurity, and use this evaluation to select models for Bot-Match.  We then conduct a visual validation of the selected models using data associated with the 2018 US Midterm elections. Finally, we describe where Bot-Match fits in the social cybersecurity workflow and illustrate the use of the Bot-Match methodology in detecting disinformation actors in the 2019 Canadian national elections.

\section{Review of Past Work and Evaluated Models}

This paper merges concepts from information retrieval, network embedding, document embedding, semi-supervised learning, link prediction, social recommendation, nearest neighbors classification, and social media analytics.  While we don't have time to go into depth in each of these deep fields of study, we will highligh relevant and related research below. 

\subsection{Content Based Information Retrieval}

Within information retrieval, the idea of object reverse search is called Content Based Information Retrieval (CBIR), and has primarily been applied to images and other multi-media \cite{lew2006content}.  Allegedly, the idea of CBIR was born in a 1992 National Science Foundation Workshop \cite{thompson2017picture}.  Information professionals often use CBIR for 1) discovering digital content on the web and 2) understanding how images are reused \cite{thompson2017picture}. Few CBIR examples exist outside of multi-media search and analysis.  These include spatial and textual search \cite{lu2011reverse} which searches based on spatial and textual similarity. 

Query on social networks is called Social Information Retrieval (SIR) \cite{bouadjenek2016social,goh2008social}.  Research in SIR has focused on social web search (improving classic information retrieval with social data), social search (sourcing questions to a social network), and social recommendation (using social networks to improve item recommendation) \cite{bouadjenek2016social}. These have led to the emergence of social content search systems like \emph{TwitterSearch}, \emph{Social Bing}, and others.  Social search is largely focused on how to leverage the rich tags that are used on many social platforms.  Social Recommendation research focuses on classic collaborative filtering tasks.  For example, SocialGCN \cite{wu2018socialgcn} was introduced in 2019 as a way to improve classic collaborative filtering with Graph Convolutional Networks to improve recommending items to users.   Similarly, Zhang et al. \cite{zhang2020social} uses matrix factorization with social regularization and the user scoring matrix to improve social recommendation. 

Related to our research, Choumane \cite{choumane2014semantic} proposed social information retrieval based on semantic similarity.  Weng et al. \cite{weng2014privacy} demonstrated a privacy-preserving search for content using rich data that had been hashed.  Privacy of both the query and the data store are preserved in this way.  Neither of these specifically focused on embedding accounts for social cybersecurity

\subsection{Embedding}

An embedding is a structure preserving map of one mathematical structure into another.  The mathematical structure of X mapped into Y is defined as $f \: X \hookrightarrow Y$.  In our case we intend to map semantic structure and graph structure into euclidean space, both separately and then simultaneously.  The embedding of semantic space has been an active research area since the 1950's \cite{harris1951methods} and graph embedding dates back to at least the 1960's \cite{tutte1960convex} with combinatorial approaches.  In this section we will highlight past work and motivate our selection of evaluated models for semantic and graph embedding.

All of these models are transductive.  The learned embedding allows us to find new data that is already represented within the network, but does not allow us to label new data from a different graph.  

\subsection{Semantic Embedding}

\noindent Given the success of word embeddings \cite{mikolov2013efficient}, researchers have attempted to develop embeddings for phrases, sentences, and even documents.  Early approaches simply averaged word embeddings for sentences and documents \cite{pennington2014glove} which were expanded in sent2vec \cite{pagliardini2017unsupervised} using word and ngram embeddings while simultaneously training the composition and the embedding.  Doc2vec \cite{le2014distributed} extended the concept of word2vec to paragraphs and other variable length text.  To do this, doc2vec introduced a fixed length paragraph vector to represent variable length text that is trained by predicting words in the paragraph.  Other approaches use a Recursive Neural Net (RNN) approach as demonstrated by the Skip-Thought model \cite{kiros2015skip}.   These were later trained over Natural Language Inference (NLI) to achieve improved results \cite{triantafillou2016towards}.

These approaches are often developed for a single language at a time, and are therefore limited on Social Media where most large conversations have multiple languages represented.  Multi-lingual embeddings have been accomplished by learning jointly on parallel corpora \cite{ruder2019survey} or by training independently and then mapping to a shared space with a bilingual dictionary \cite{artetxe2018robust}.  Two competing models for universal encoding are 1) Google's Universal Sentence Encoder \cite{cer2018universal} and Facebook's Language-Agnostic SEntence Representations (LASER) toolkit \cite{artetxe2018massively}.  The Google Universal Sentence Encoder (USE) encodes sentences and short paragraphs using two models, the Transformer model and the Deep Averaging Network (DAN) model. The Transformer model uses the encoding subgraph of the transformer architecture to create context aware embedding.  The DAN model uses a feed forward deep neural network to average word and bigram representations.  Facebook's Laser toolkit uses an encoder/decoder approach with Bidirectional Long Short-term Memory (BiLSTM) trained on 223 million sentences to create a universal encoding scheme for 93 languages.  In our implementation Google USE was trained on cleaned and concatenated user content and Facebook Laser was trained at the individual tweet level a then tweet level embeddings were averaged to create a user/node embedding.

Prior to word and sentence embedding, researchers attempted to analyze topic groups with several methods, notably Latent Dirichlet Allocation \cite{blei2003latent} and Latent Semantic Analysis/Indexing \cite{deerwester1990indexing}.  Both Latent Direchlet Allocation (LDA) and Latent Semantic Analysis (LSA) are used to discover latent topics found in a corpus of documents and to reduce dimensionality.  In the course of assigning documents to a fixed number of topics, both models are also reducing the dimensions of the corpus and inherently creating a document embedding.  These approaches operate on a bag of words or tokens and are inherently multi-lingual (assuming appropriate language parsers).  Latent Dirichlet Allocation (LDA) uses a probabilistic statistical model to map documents to topics while Latent Semantic Analysis/Indexing (LSA/LSI) uses singular value decomposition to reduce the dimensions and thereby producing a set of topics or concepts.  We used term frequency-inverse document frequency (TF-IDF) for LSA, but term frequncy for LDA since Blei describes how LDA was created to overcome some of the shortcomings of TF-IDF.

From the discipline of collaborative filtering we find another approach of measuring similarity and delineating neighbors without creating an embedding.  Starbird, Muzny and Palen used collaborative filtering to find social media users on the ground during natural disasters \cite{starbird2012learning}.  Memory based collaborative filtering employs similarity measures to identify neighbors and thereby make recommendations for item-users data \cite{sarwar2001item}.  Using a similar approach we leverage Cosine and Jaccard similarity to measure similarity between user's content based on bag of words representation.  Given a bag of words representation, cosine similarity measures the cosine of the angle between term frequency-inverse document frequency (TF-IDF) vector representations of the document.  Jaccard similarity, on the other hand, compares the relative intersection of two documents.  Cosine similarity performs better on TF-IDF, whereas Jaccard similarity performs better on a Bag-of-Words representation.  Cosine similarity will take into consideration frequencies, Jaccard similarity will only consider the presence or absence of words.  This offers a baseline for comparison of more complex methods, and as we discover performs surprisingly well  when accounts have produced substantial content.

\begin{table*}[htbp]
  \centering
  \caption{Model Description by Type}
  \resizebox{\linewidth}{!}{%
    \begin{tabular}{cclll}
    Type  & Subtype & \multicolumn{1}{c}{Model} & \multicolumn{1}{c}{Data} & \multicolumn{1}{c}{Embed Dim} \\
    \midrule \midrule
    \multirow{6}[6]{*}{Content} & \multirow{2}[2]{*}{Collaborative Filtering} & Jaccard Similarity & Term Frequency & No embedding \\
          &       & Cosine Similarity & Term Frequency & No embedding \\
\cmidrule{2-5}          & \multirow{2}[2]{*}{Topic Modeling} & LDA   & Term Frequency & N x 200 \\
          &       & LSI   & TFIDF & N x 200 \\
\cmidrule{2-5}
          &   Document    & Doc2Vec & User Text & N x 200 \\
\cmidrule{2-5}          & \multirow{2}[2]{*}{Universal } & Google USE & User Text & N x 512 \\
          &       & Facebook LASER & Tweet Text & N x 1024 \\
    \midrule
    \multirow{8}[6]{*}{Network} & \multirow{3}[2]{*}{Factorization} & Graph Factorization & Adjacency Matrix & N x 32 \\
          &       & HOPE  & Adjacency Matrix & N x 128 \\
          &       & BigGraph & Edge list & N x 1024 \\
\cmidrule{2-5}          & \multirow{4}[2]{*}{Random Walk} & node2vec & Edge list    & N x 64 \\
          &       & Splitter & Edge list & N x 128 \\
          &       & role2vec & Edge list & N x 128 \\
            &       & SybilRank & Adjacency List & No embedding \\
\cmidrule{2-5}          & \multirow{2}[2]{*}{Deep Learning} & SDNE  &   Adjacency Matrix    & N x 128 \\
          &       & GCN (no features) & Adjacency Matrix  & N x 32 \\
    \midrule
    \multirow{3}[4]{1.2cm}{\centering Network \& Content} & \multirow{2}[2]{*}{Deep Learning} & GCN with Features & Adjacency Matrix   & N x 32 \\
    &  & GraphSAGE  & Adjacency Matrix \& BoW  & N x 50 \\
\cmidrule{2-5}          & Factorization & BigGraph with initial  & Edge list w/ Embedding & N x 1024 \\
    \bottomrule
    \end{tabular}}%
  \label{tab:addlabel}%
\end{table*}%

\subsection{Graph Embedding}

While most graph based analysis is designed to operate on the original adjacency matrix or equivalent structure, recently methods have been devised to embed the graph in vector space.  Vector space representations of graphs have applications in node classification, link prediction, clustering, and visualization \cite{goyal2018graph}.  In this paper graph embedding is specifically focused on embedding nodes into vector space, not embedding the entire graph in vector space.

For the purposes of this research we've adopted the topology that Goyal and Ferrara introduced \cite{goyal2018graph}.  They divide graph embedding techniques into methods based on 1) Factorization, 2) Random Walk, and 3) Deep Learning.  In our research we will test prominent models from each of these categories.

Factorization methods use various methods to factorize the adjacency matrix or other matrix representing the graph (Laplacian matrix, Katz similarity matrix, others).  Eigenvalue decomposition can be used on matrices that are positive semi-definite, otherwise gradient descent methods are used.  The primary factorization models we tested were the High-Order Proximity preserved Embedding (HOPE) algorithm \cite{ou2016asymmetric} and Facebook Biggraph \cite{lerer2019pytorch}, with Singular Value Decomposition of the Adjacency Matrix used as a baseline.  HOPE preserves higher order proximity by minimizing $ \left| \left| S - Y_S Y_T \right| \right| ^ 2 $  where S is a similarity matrix, instead of the adjacency matrix.  In our case we used the Katz index to create the similarity matrix.  Katz centrality is defined as

\[
C_{katz}= \sum_{k=1}^{\infty}\sum_{j =1}^{n}\alpha ^ k \left( A^k \right)_{jk}
\]
where $A$ is the adjacency matrix and $\alpha$ is the attenuation factor (smaller than the absolute value of the largest eigenvalue of A).

In addition to using the HOPE algorithm, we also tested Facebook's Pytorch-Biggraph toolbox \cite{lerer2019pytorch}.  Pytorch-Biggraph can embed large graphs using several available factorization based models (TransE, RESCAL, DistMult and ComplEx).  Pytorch-Biggraph overcomes complexity and memory constraints by partitioning the graph and then using multi-threaded and distributed execution with batched negative sampling.  In testing Pytorch-Biggraph we wanted to determine what our loss of performance would be for scalability.  We tested three settings for the BigGraph algorithm.  The first setting is with a single edge type as is assumed in all other models.  BigGraph allows the user to define different edge types, which was also tested and reported using our three edge types (\emph{mention}, \emph{retweet}, and \emph{reply}). This multi-modal approach did not perform well, and is not included in our results. The third and final setting that we test and report is BigGraph with an initial value, in our case a content embedding produced by LDA.  This would initialize the training with knowledge learned from the content similarity, and as reported below often improved BigGraph performance (though in this setting BigGraph requires an initialization computation which may be computationally costly and which makes BigGraph no longer an end-to-end solution).

Multiple models leverage random walks to embed a graph.  The model that we used is the \emph{node2vec} model \cite{grover2016node2vec} which uses a biased random walk procedure to explore neighborhoods while preserving higher order proximity between nodes. We also tested the Splitter \cite{epasto2019single} random walk based algorithm that is tailored for social networks and embeds multiple persona based representations of each user and then combines these to produce a single embedding for the user.

Also within the random walk family of models we implement \emph{role2vec} \cite{ahmed2018learning}.  The \emph{role2vec} algorithm leverages attributed walks which maps an input vector to a vertex type.  The embedding structure for \emph{role2vec} differs from all of our other methods in that it embeds the vertex type and not necessarily the vertex neighborhood.  Our implementation uses the Weisfeiler-Lehman kernel \cite{shervashidze2011weisfeiler} to extract vertex features.  SybilRank \cite{cao2012aiding} is based on random walk but does not produce an embedding and is discussed in more detail below.

Our primary Deep Learning model was the Structural Deep Network Embedding (SDNE) model \cite{wang2016structural} that uses deep autoencoders and decoders to preserve 1st and 2nd order network proximities.  This is accomplished by optimizing both proximities simultaneously.  The decoder is based on Laplacian Eigenmaps that penalizes similar vertices that are embedded far apart.

\subsection{Content and Network Embedding}

By its very nature social media data contains rich node features to include rich semantic features.  Graph convolutional networks \cite{kipf2016semi} have emerged as a way to embed a network simultaneously with the respective node features.  While a variety of node feature representations exist in social media, we primarily used GCN's to simultaneously encode the network while considering node content features (the combined tweet text produced by each account).
GCN scales better than SDNE by iteratively applying a convolution operator on the graph and aggregating the embedding of neighbors.  The GCN defines a function of the form
\[
f(X,A) = softmax \left( \hat A \:  ReLU \left ( \hat A \: XW^{(0)} \right)W^{(1)} \right)
\]
where X represents node features and A represents network adjacency matrix, $\hat A = D^{\frac{1}{2}} (A + I_N) D^{\frac{1}{2}}$. In addition to Kipf's original GCN autoencoder, we will also evaluate GraphSAGE (SAmple and AggreGatE) \cite{hamilton2017inductive}, which extends GCN's to generate inductive embeddings by sampling and aggregating each node's neighborhood features.  GraphSAGE is designed to embed larger networks than the original GCN as we'll demonstrate below.

\subsection{Similarity Based Approaches to Bot Detection}

Several past research efforts are somewhat related to our use of similarity measures to detect bots.  One well-known and often cited unsupervised machine learning tool is the Debot algorithm \cite{chavoshi2016debot} which uses warped correlation to find correlated accounts.  These correlated accounts are bot accounts that post the same content at roughly the same time.  Also notable is a model by Xiao et al \cite{xiao2015detecting} that uses various features to classify entire clusters of new accounts on LinkedIn to detect batches of fake accounts.  Magelinski et al \cite{magelinski2019graph} demonstrates bot detection with graph classification by extracting a graph's latent local features and binning nodes together along 1-D cross sections of the feature space.

Finally, Ali Alhosseini et al. \cite{ali2019detect} uses GCN's with network, content, profile, and propagation features to train a supervised machine learning model to detect fake news given URL and cascade-wise detection.  This study also explains and measures the decrease in bot detection performance as a model ages.  Note that this research explicitly uses the GCN and geometric deep learning in a supervised manner, and while visualizing lower dimension representation of bot detection features, does not use these models in an unsupervised or semi-supervised manner as proposed by Bot-Match. 

\subsection{Other Related Approaches}

\subsubsection{Semi-supervised Label Propagation Approaches}

Research has used semi-supervised methods to detect social bots (sometimes called sybils) since the earliest days of social networks.  These methods include representative (linearlized) label propagation methods \cite{gong2014sybilbelief,wang2017gang}, random walk methods \cite{yu2006sybilguard,jia2017random}, and recently  adapted Graph Convolutional Network (GCN) methods for sybil detection with a few labeled instances \cite{li2018deeper}. 
There are several key differences between our method and approach and that of these semi-supervised models.  First, many of these assume that the network contains a sybil (bot) region and a benign region, with attack edges between them.  We have found that this is not always the case, and many times the most sophisticated bots are tightly embedded in the target audience (assumed to be benign).  Second, while many of these methods use very few labels, most of the papers evaluate their methods with a sample of 100 \cite{jia2017random} to several thousand instances \cite{wang2017gang}.  In our case, we often only have one instance to query with.  Finally, we want to return a ranked list not a classification.  Label propagation can at times return a ranked list and has seen limited use for information retrieval \cite{yang2006document}.  For example, SybilRank \cite{cao2012aiding} can return a ranked list and has been used as a semi-supervised model to rank possible Sybil accounts.  SybilRank does assume a Sybil region and benign region of the network, with attack edges between them (in other words there is tight clustering among sybils and human users).  Nonetheless, we include SybilRank for comparison purposes in our evaluation.  SybilRank (and other label propagation methods) do not produce an embedding, and can therefore have faster run times if a user only performs a handful queries.  Additionally, unlike some of our methods, SybilRank easily resolves queries with more than one account.  

\subsubsection{Link Prediction}

In our task of measuring similarity between users in an online social network, we will borrow some concepts from social link prediction.  For example, \cite{barbieri2014follow} proposes a social user recommendation method that recommends based on either `topic' or `social' link.  This is similar to our desire to find similarity semantically and socially.  Our method must diverge from this in that existing links must be considered.

\section{Data}

\subsection{Building Networks and Cleaning Text}

Bot-Match assumes that the user has a filtered stream or conversation that she is trying to analyze for the presence of disinformation.  This could be an online discussion around a topic (i.e. climate change), a political event (Canadian 2019 Elections ), or a natural disaster (Hurricane Harvey).  Note that in all cases the individual tweets are part of a larger connected online conversation and are not randomly selected from the social media environment (network analysis requires a network).  

Bot-Match also assumes some level of network homophily as well as a static network snapshot.   We assume network links are far more likely to happen with accounts that are similar or have a similar narrative than with accounts that are different.  While some communication links may confront an account that is different, we assume the majority of links are with similar accounts.  In social networks this is often referred to as homophily and defined as ``birds of a feather flock together'' \cite{mcpherson2001birds}. Additionally, we assume that the analyst is focused on a snapshot of a network (following links at time x or communication links during the month of January).  While some of the methods discussed below could adapt to dynamic networks, many of them could not.

\begin{figure}[htb]
\centering
  \includegraphics[width=0.7\columnwidth, trim={3cm 1cm 3cm 1cm},clip]{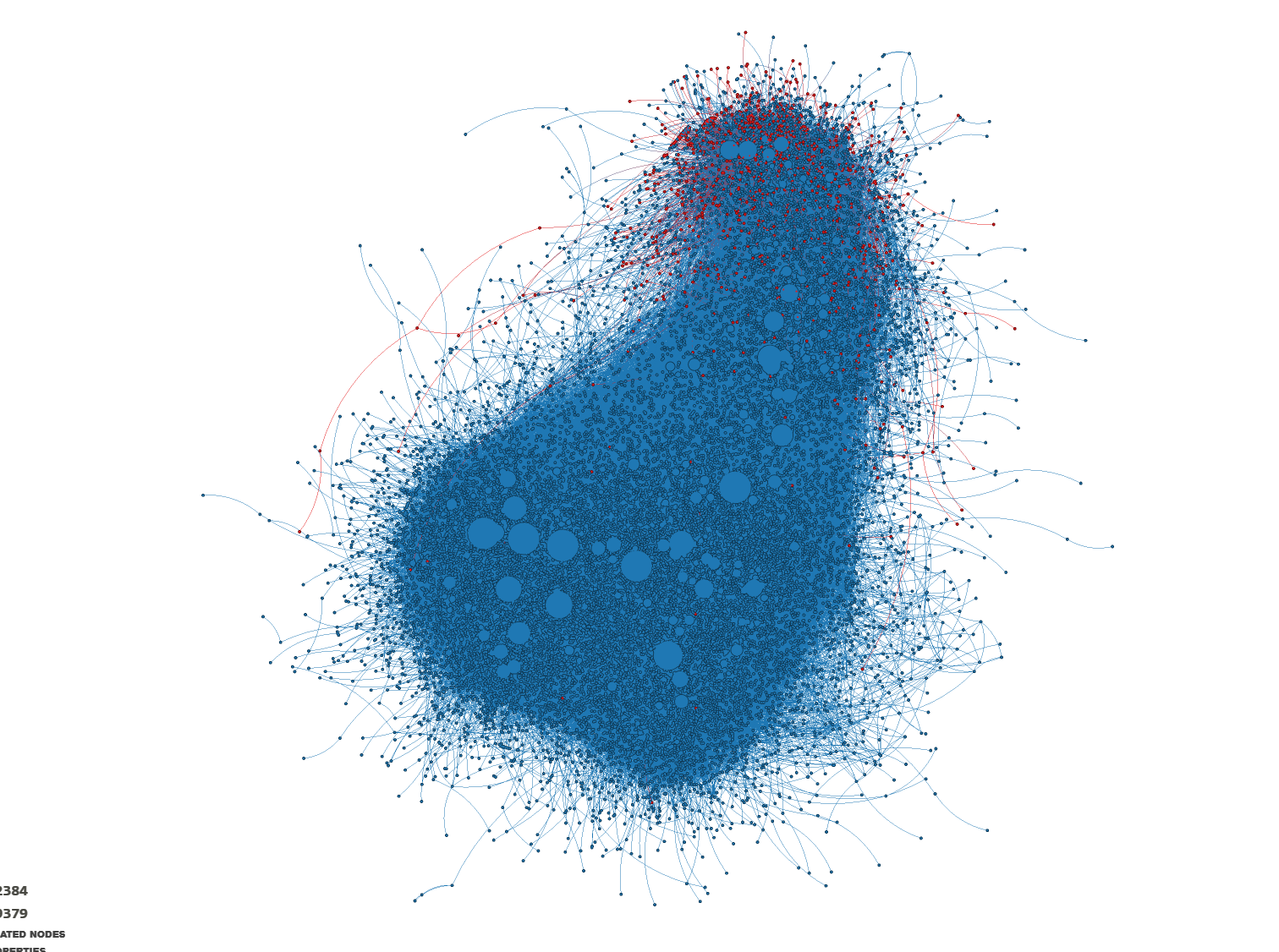}
  \caption{Conversational Network of Followers of a Journalist in Yemen.  Red denotes random string accounts that were part of an intimidation campaign.  Network includes 35,763 nodes and 195,172 edges.}
  \label{fig:ionanetwork}
\end{figure}

Below we list two data sets that we collected in order to evaluate Bot-Match.  Bot-Match is designed to find similar bot accounts that have evaded a supervised machine learning initial approach.  In order to test Bot-Match, we needed to find data sets where we can separately identify similar accounts, label them, and then test Bot-Matches ability to find them given a single seed node.  The two data sets selected are discussed below:

\subsection{Yemen Data}

The first data set is the combined tweets produced by all followers of a freelance journalist in Yemen.  Starting in the Fall of 2017, a determined and documented intimidation attack was launched against her Twitter Account \cite{yemen}.  The intimidation attack was characterized by a surge of strange accounts, many of them with strange and disturbing images or threatening messages. Many of these intimidation accounts were distinguished by 15 digit randomly generated alpha-numeric strings for their screen name, such as \textbf{gG6RKc6QBqOLKyU} (not real screen name).  We developed a logistic regression classifier to detect random strings based on features consisting of character n-grams and string entropy.  Using this model we were able to achieve 94.25\% accuracy in identifying random string accounts in Twitter, allowing us to automatically label 4,312 accounts characterized by a random string screen name and likely part of the coordinated intimidation campaign.  While these accounts do not compose the entire intimidation attack, they nonetheless present an interesting social cybersecurity account that we can externally label and test Bot-Match performance in finding them given a query node.  In our experiment, we will test and see if nearest neighbor searches of various embeddings would be able to find these random string intimidation accounts if we were not able to label them by their screen name.  Throughout the rest of the paper this data will be called the \emph{Yemen} data. The conversational network of the Yemen data is provided in Figure \ref{fig:ionanetwork} with random string intimidation accounts colored red.  Data description is provided in Table \ref{tab:data_table}.  

\subsection{Internet Research Agency Data}

The second data set consists of tweets produced by Russia's Internet Research Agency around the time of the 2016 US Elections and released by Twitter in October 2018 \cite{ira_data}.  The St Petersburgh based Internet Research Agency (IRA) is a company that conducts focused online information operations on behalf of the Russian government and Russian businesses.  The IRA represents one of the more experienced organizations involved in state-sponsored disinformation \cite{beskowIRA2020}.  Twitter detected deliberate manipulation by the IRA, suspended the accounts and released the related data in an elections transparency effort (similar manipulation has been associated with Iran, Venezuela, China, and Spain and related data released).  

\begin{figure}[htb]
\centering
  \includegraphics[width=0.7\columnwidth, trim={0cm 0cm 0cm 0cm},clip]{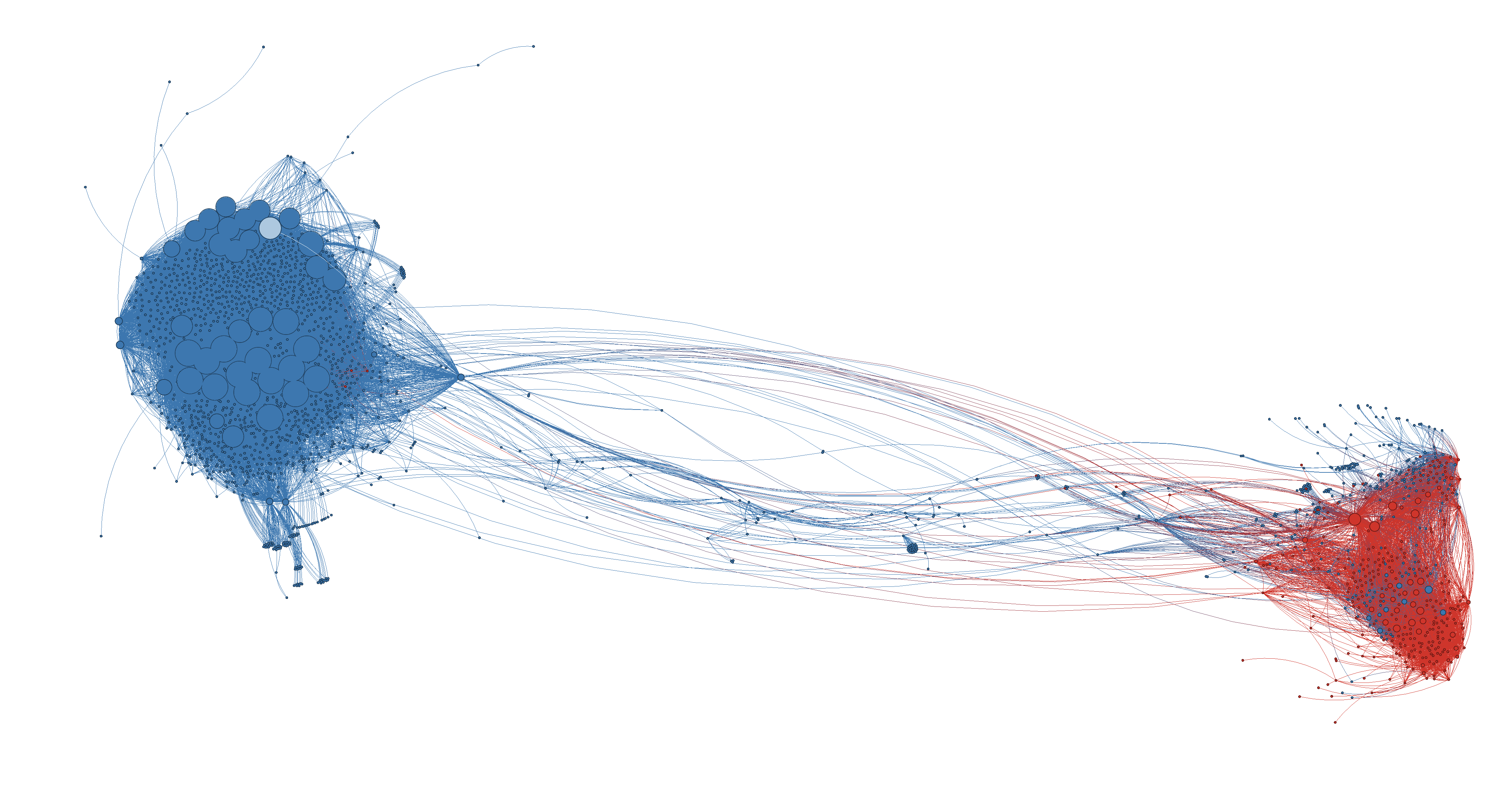}
  \caption{Conversational Network of Russian Internet Research Agency data released by Twitter.  Red denotes accounts that targeted African American communities.  Network includes 1,958 nodes and 35,931 edges.}
  \label{fig:iraanetwork}
\end{figure}

The data demonstrates that the IRA specifically targeted African American online users in an effort to increase racial tensions in the United States \cite{black_nyt}.  For the purposes of testing Bot-Match we will label any account that shared relevant hashtags targeting African American populations as an account that is participating in this effort.  In our case study we will test the performance of Bot-Match to detect these accounts after removing all hashtags.  The conversational network of the IRA data (removing all nodes that didn't produce tweets in the released dataset) is provided in Figure \ref{fig:iraanetwork}. Data description is provided in Table \ref{tab:data_table}.

\begin{table}[htbp]
  \centering
  \caption{Data Summary}
    \begin{tabular}{lcccc}
    \multicolumn{1}{c}{Data} & Yemen Data & IRA Data & CORA Data & Reddit Data \bigstrut[b]\\
    \hline
    Users/Nodes & 35,763 & 1,958 & 2,708 & 231,443\bigstrut[t]\\
    Total Edges & 195,172 & 35,931 & 5,278 & 11,606,919\\
    Tweets & 4,535,117 & 9,041,308 &  &  \\
    Top Languages & en,ar,fr,es & ru,en,de,uk & & \\
    Retweet Edges & 108,382 & 31,398 & & \\
    Mention Edges & 50,933 & 859 & & \\
    Reply Edges & 35,857 & 4,122 & & \bigstrut[b] \\

    \hline
    \end{tabular}%
  \label{tab:data_table}%
\end{table}%

\subsection{Benchmark Data}

In addition to the two data sets evaluated above, we also included two bench mark data sets.  For evaluation we require network data with semantic node features and node labels.  We chose to use the Cora citation dataset which contains a bag of words (BoW) for semantic node features.  Cora's network topology and properties are substantially different than online social networks, but it was selected due to its frequent use in network related research.  We also chose to use a larger social network benchmark dataset derived from Reddit which uses semantic embedding features for node features.  This was created and used by Hamilton et al. to evaluate GraphSAGE \cite{hamilton2017inductive}.  

\subsection{Data Processing}

Given curated, filtered, and related social media data, Bot-Match first builds the communication network edgelist by assigning a source and target for the directional communication. In the Twitter environment, this means creating directed links between accounts that \emph{mention}, \emph{retweet}, or \emph{reply} to each other.  Once the network is created, we also remove any nodes that are not in the dataset (meaning we don't have content and node features for them).  For example, a user may be mentioned in the data set but they never produced a tweet that ended up in the dataset.  These users are therefore removed, as well as any isolates that remain.  If keeping the users was required then collecting their content/timeline would be an additional data requirement. 

For most of our models, \emph{retweet}, \emph{reply}, and \emph{mention} edges are treated equally as directed communication links.   Only Facebook Big Graph algorithm will take into consideration the categories of links, with mixed results.  In practice the analyst could choose to embed a single modality of interest, for example only the retweet network.

Unless otherwise noted (see comments on Facebook LASER), the text associated with each user (node) is a concatenation of all social media posts associated with that user.  To clean the text we removed URLs, punctuation, reserved words, emojis and smileys.  We removed hashtags and mentions with the IRA data but left them in for the Yemen data.  They were removed from the IRA data since they were used to label the data and would artificially inflate content embedding models.

\section{ Model Evaluation and Validation}

To evaluate embedding models for the Bot-Match methodology with the Yemen and IRA data, we created the respective \emph{content}, \emph{network}, and \emph{network + content} embeddings for each data set, and then used these embedding to search for nearest neighbors of positively labeled accounts.  For each data/embedding combination, we calculated the k nearest neighbors for $k \in \{ 10, 50 \}$ and then measured the precision of the response, defined as the proportion of positive responses.  After conducting this query with each positively labeled account as the seed node, we averaged the precision across all queries to compute a metric for the given data/embedding combination.  

Given that we calculated k nearest neighbors for $k \in \{ 10, 50 \}$, our primary metric was precision at $k = 10$ (p@10) and precision at $k = 50$ (p@50).  Given that the labels aren't ranked, we cannot leverage any rank based metric, and precision at n is therefore appropriate.  We can compare these percentages to the naive approach of random sampling, which is provided in Table \ref{tab:emb_results}.  Any performance over these random values indicates model value. 

In our evaluation, each query is an independent test, and is not meant to simulate an analyst recursively searching the entire network.  In social cybersecurity application users generally recursively use Bot-Match two or three `hops' from the seed node.  We've never seen users try to traverse an entire network given the unwieldy nature of this task.  Because of this, our evaluation and the production Bot-Match methodology are not meant to estimate the total number of bots in a network (supervised machine learning models or label propagation models are better suited for this task).  Without estimating the total number of bots in a stream, evaluating the recall and F1 score are not appropriate.

The results for the embedding test are found in Table \ref{tab:emb_results} and provide insights on both specific models and appropriate use cases for model \emph{types}.  The first observation is that all models provide significant value to the user when compared to naive baseliness.  For example, in the case of the Yeman data, if a user queries with a single random string intimidation account, on average 5 out of 10 returned accounts will be random string intimidation accounts (and the remainder may be other types of intimidation accounts used in this attack).  This provides real and tangible value to analysts.  The second observation is that all models perform better on IRA data than Yemen data.  This is because the IRA data is smaller with a higher density of ``similar'' accounts and is more easily distinguished in both graph and semantic representation.  

\begin{table}[htbp]
\resizebox{\linewidth}{!}{%
   \begin{threeparttable}
  \centering
  \caption{Model Evaluation}
   
    \begin{tabular}{clcc @{\hskip 0.3in} cc @{\hskip 0.3in} cc @{\hskip 0.3in} cc}
          &       & \multicolumn{2}{c}{Yemen Data} & \multicolumn{2}{c}{IRA Data} & \multicolumn{2}{c}{CORA} & \multicolumn{2}{c}{Reddit} \\
\cline{3-10}    Type  & \multicolumn{1}{c}{Model} & p@10 & p@50 & p@10  & p@50  & p@10  & p@50  & p@10  & p@50 \bigstrut[t] \bigstrut[b] \\
    \hline
          & Random Baseline & 0.12 & 0.12 & 0.22  & 0.22  & 0.18  & 0.18  & 0.045 & 0.045 \bigstrut[t] \bigstrut[b]\\ \hline
    \multirow{7}[2]{*}{Content} & Jaccard Similarity & 0.421 & 0.387 & \textbf{0.868} & \textbf{0.854} & \textbf{0.604} & 0.457 & *    & * \bigstrut[t]\\
          & Cosine Similarity & 0.371 & 0.303 & 0.835 & 0.808 & 0.600 & \textbf{0.458} & 0.371 & 0.255 \\
          & LDA   & 0.457 & 0.378 & 0.776 & 0.711 & 0.523 & 0.450 & *    & * \\
          & LSA   & 0.424 & 0.359 & 0.796 & 0.780 & 0.484 & 0.339 & \textbf{0.474} & \textbf{0.362} \\
          & Doc2Vec & \textbf{0.461} & \textbf{0.410} & 0.808 & 0.736 & *     & *     & *    & * \\
          & Google USE & 0.426 & 0.373 & 0.716 & 0.681 & *     & *     & *    & * \\
          & Facebook LASER & 0.454 & 0.406 & 0.757 & 0.616 & *     & *     & *    & * \bigstrut[b]\\
    \hline
    \multirow{9}[2]{*}{Network} & SybilRank & \textbf{0.510} & 0.174 & 0.675 & 0.523 & 0.730 & 0.669 & 0.720    & 0.724 \bigstrut[t]\\
          & Graph Factorization & 0.391 & 0.258 & 0.625 & 0.55  & 0.555 & 0.451 & 0.796 & 0.77 \\
          & HOPE  & 0.342 & 0.283 & 0.715 & 0.54  & 0.680 & 0.539 & **    & ** \\
          & BigGraph  & 0.335 & 0.237 & 0.734 & 0.65  & 0.504 & 0.302 & 0.813 & 0.641 \\
          & node2vec & 0.365 & \textbf{0.309} & 0.734 & 0.66  & 0.704 & 0.456 & \textbf{0.921} & \textbf{0.907} \\
          & Splitter & 0.258 & 0.172 & 0.663 & 0.605 & 0.335 & 0.238 & **    & ** \\
          & role2vec & 0.326 & 0.248 & \textbf{0.772} & \textbf{0.75} & \textbf{0.768} & \textbf{0.651} & **    & ** \\
          & SDNE  & 0.396 & 0.303 & 0.701 & 0.61  & 0.452 & 0.376 & **    & ** \\
          & GCN (no features) & 0.285 & 0.213 & 0.612 & 0.58  & 0.566 & 0.339 & **    & ** \bigstrut[b]\\
    \hline
    \multicolumn{1}{c}{\multirow{3}[2]{1.2cm}{\centering Network \& Content}} & GCN with Features & \textbf{0.459} & \textbf{0.397} & 0.685 & 0.632  & 0.756 & 0.630  & **    & **  \bigstrut[t]\\
          & GraphSage & 0.436 & 0.382 & 0.644 & 0.565  & \textbf{0.806} & \textbf{0.733} & \textbf{0.857} & \textbf{0.840} \\
          & BigGraph w/ Initial & 0.356 & 0.250 & \textbf{0.761} & \textbf{0.695} & 0.516 & 0.305 & 0.806 & 0.612 \bigstrut[b]\\
  \hline
    \end{tabular}%
    \label{tab:emb_results}%
      \begin{tablenotes}
      \small
      \item * These models are not able to embed the provided data structure
      \item ** These models were not able to embed large graphs on our compute resources
    \end{tablenotes}
    \end{threeparttable}}
  
\end{table}%

Next we'll compare the embedding types.  With the much more integrated conversation found in the Yemen data, we see the \emph{content} models generally provide better precision across all values, while the more clustered IRA data has almost equal performance by both \emph{content} and \emph{network embedding}.  The \emph{network + content} embedding provides the best model for the Yemen data, and still outperforms most network models for the IRA data, albeit with a Biggraph as opposed to GCN.  

Focusing on the \emph{content} algorithms, we see the classic models excel in similarity analysis.  While these models will not necessarily create the contextual universal embedding that Google USE and Facebook LASER were designed for, they still excel at the basic task of document similarity and document retrieval.   In particular the LDA model and Jaccard similarity perform exceptionally well, are inherently multi-lingual based on a bag of words or bag of tokens (though care must be taken when choosing the size of term frequency matrix in the presence of many languages).   We also observe that while Doc2Vec has high performance, it only marginally surpasses LDA.  The strong performance by Cosine/Jaccard similarity and LDA is largely a result of having substantial content for each user.  The value of a Bag of Words representation increases with more content.

Focusing on graph embedding, we see strong performance by random walk algorithms \emph{node2vec}, \emph{role2vec}, and \emph{SybilRank} across all datasets.  Graph factorization and deep learning showed some success, though this fades at higher levels of $n$.  The Pytorch Biggraph model scales much better than any other model, but did not perform as well as other models in our implementation.  The GCN model without features and the Splitter model did not perform well in our evaluation. SybilRank performs strong across the data sets at lower values of $k$, though performance at higher $k$ varied.

Combining graph embedding with node features produced strong but mixed results.  While GCN's produced the highest performance on the Yemen data, it produced mediocre performance on IRA data, with the reverse true for BigGraph (high results for IRA but less so for Yemen).  This demonstrates that on this social data that the GCN is getting more traction on the content features as opposed to the BigGraph algorithms which is primarily focused on network features.  GraphSAGE demonstrated strong performance and scaled better than GCN.  BigGraph scaled better than all other models.  

Given these results we selected Bot-Match algorithms for social media embedding in a social cybersecurity context.  For our production Bot-Match algorithm, we offer LDA for content similarity, node2vec for network similarity, GCN with features for Network and Content embedding.  We also use Cosine Similarity and BigGraph implementations available for larger networks ($>100K$ nodes).

\begin{figure*}[htb]
\centering
\begin{subfigure}{.23\linewidth}
  \centering
  \includegraphics[width=\linewidth, trim = {0 0cm 0 0cm}, clip]{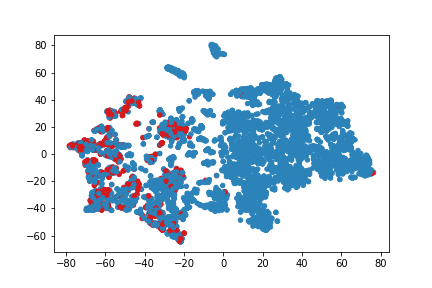}
  \caption{Yemen LDA}
  \label{fig:bridge_sub1}
\end{subfigure}%
\begin{subfigure}{.23\linewidth}
  \centering
  \includegraphics[width=\linewidth, trim = {0 0cm 0 0cm}, clip]{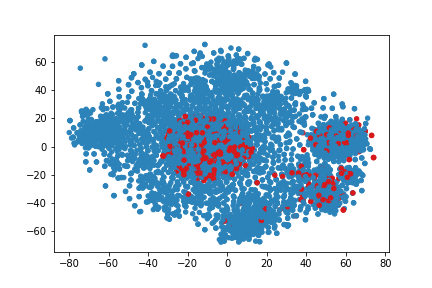}
  \caption{Yemen Node2vec}
  \label{fig:bridge_sub2}
\end{subfigure}
\begin{subfigure}{.23\linewidth}
  \centering
  \includegraphics[width=\linewidth, trim = {0 0cm 0 0cm}, clip]{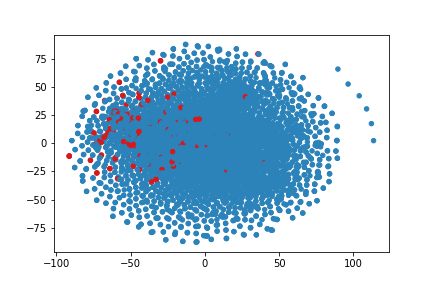}
  \caption{Yemen BigGraph}
  \label{fig:bridge_sub3}
\end{subfigure}
\begin{subfigure}{.23\linewidth}
  \centering
  \includegraphics[width=\linewidth, trim = {0 0cm 0 0cm}, clip]{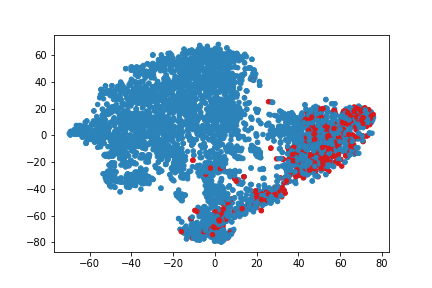}
  \caption{Yemen GCN (w/ Features)}
  \label{fig:bridge_sub4}
\end{subfigure}
\begin{subfigure}{.23\linewidth}
  \centering
  \includegraphics[width=\linewidth, trim = {0 0cm 0 0cm}, clip]{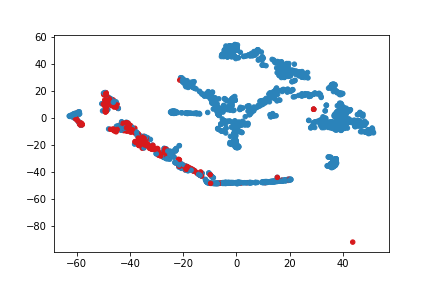}
  \caption{IRA LDA}
  \label{fig:bridge_sub1}
\end{subfigure}%
\begin{subfigure}{.23\linewidth}
  \centering
  \includegraphics[width=\linewidth, trim = {0 0cm 0 0cm}, clip]{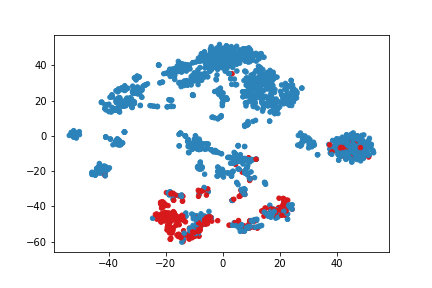}
  \caption{IRA Node2vec}
  \label{fig:bridge_sub2}
\end{subfigure}
\begin{subfigure}{.23\linewidth}
  \centering
  \includegraphics[width=\linewidth, trim = {0 0cm 0 0cm}, clip]{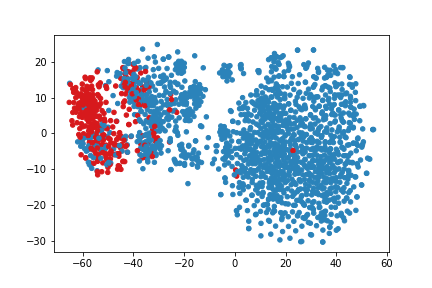}
  \caption{IRA BigGraph}
  \label{fig:bridge_sub3}
\end{subfigure}
\begin{subfigure}{.23\linewidth}
  \centering
  \includegraphics[width=\linewidth, trim = {0 0cm 0 0cm}, clip]{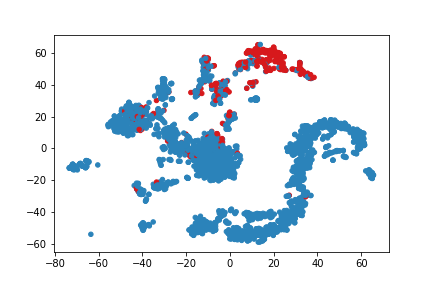}
  \caption{IRA GCN (w/ Features)}
  \label{fig:bridge_sub4}
\end{subfigure}
\caption{t-Distributed Stochastic Neighbor Embedding (TSNE) 2-D Visualization of Embeddings for both Yemen and IRA Data}
\label{fig:tsne}
\end{figure*}

In Figure \ref{fig:tsne} we provide t-Distributed Stochastic Neighbor Embedding (tSNE) visualization for the selected models for both the Yemen and the IRA data.  These visualizations provide more insight into the model performances.  We can visually see the higher precision of IRA data over Yemen data.  We can also see various natural clusters emerging from the data, particularly the graph structure already visualized for the IRA data.

\subsection{Comparison to Supervised Methods}

In this section we will compare the results seen in Table \ref{tab:emb_results} for the Yemen data set with supervised machine learning methods, namely the Bot-Hunter model \cite{beskow2018bot} and the Botometer model \cite{davis2016botornot}.  The Yemen data set was selected because it is in the data format consumed by these models.  The IRA data, while derived from Twitter data, was not released in its original format, and is therefore not compatible with the data parsers and feature engineering of these models.  

In the Yemen data set we labeled 4,312 accounts as having 15 digit randomly generated alpha-numeric strings.  These accounts were part of a larger bot intimidation attack conducted against a free lance journalist.  In the Fall of 2017, her followers jumped from approximately 20K to approximately 50K.  Most of these were bots.  At the time of our collection, we were able to get account history on 35,763 of these that had not been suspended, removed, and become a private account. 

As seen in Table \ref{tab:supervised}, Bot-Hunter and Botometer both find approximately 90\% of the 15 digit random alpha numeric screen name accounts (this is partially due to the fact that both models use screen name string entropy in their feature space).  It also displays some of the limitations of supervised learning.  Botometer, which has been tuned for high precision, only found random string bots.  100\% of the accounts found by Botometer were Random String accounts, and it was not able to identify any of the 15,000 to 20,000 additional bots.  Bot-hunter, which was tuned for higher recall did find many more non-random string bots, but 28,406 accounts is still too many to manually evaluate for characteristics and attribution.  In this case, as the user begins exploring influential bots or other suspicious accounts not found by these supervised models, Bot-Match allows them to quickly find similar accounts with content based information retrieval.  

\begin{table}[htbp]
  \centering
  \caption{Results of Supervised Machine Learning Models}
    \begin{tabular}{lcc}
          & \multicolumn{1}{l}{Total Bots} & \multicolumn{1}{l}{Percentage of Random String Accounts Found}  \\ \hline
    Bot-Hunter & 28,406 & 86.5\% \\
    Botometer & 4,312 & 89.6\% \\ \hline
    \end{tabular}%
  \label{tab:supervised}%
\end{table}%

\section{Visual Validation with 2018 US Midterm Social Media Data}

Given the four models selected above, we wanted to conduct one additional visual validation of the Bot-Match methodology in general and these four models in particular.  We had previously collected all Twitter content and connections associated with US Members of Congress or Congressional Candidates for the 2018 US Midterm elections.  We decided to use this data to test our embeddings since it is easily labeled by both party (\textit{Republican}, \textit{Democrat}, \textit{Other}) and by chamber (\textit{House} or \textit{Senate}).  We wanted to test if the Bot-Match methodology and the four models selected above could leverage the social media connections (friend connections) and the social media content to capture the complex political environment of the US bicameral legislature in Euclidean vector space.  

\begin{figure*}[bht]
\centering
\begin{subfigure}{.23\linewidth}
  \centering
  \includegraphics[width=\linewidth, trim = {0 0cm 0 0cm}, clip]{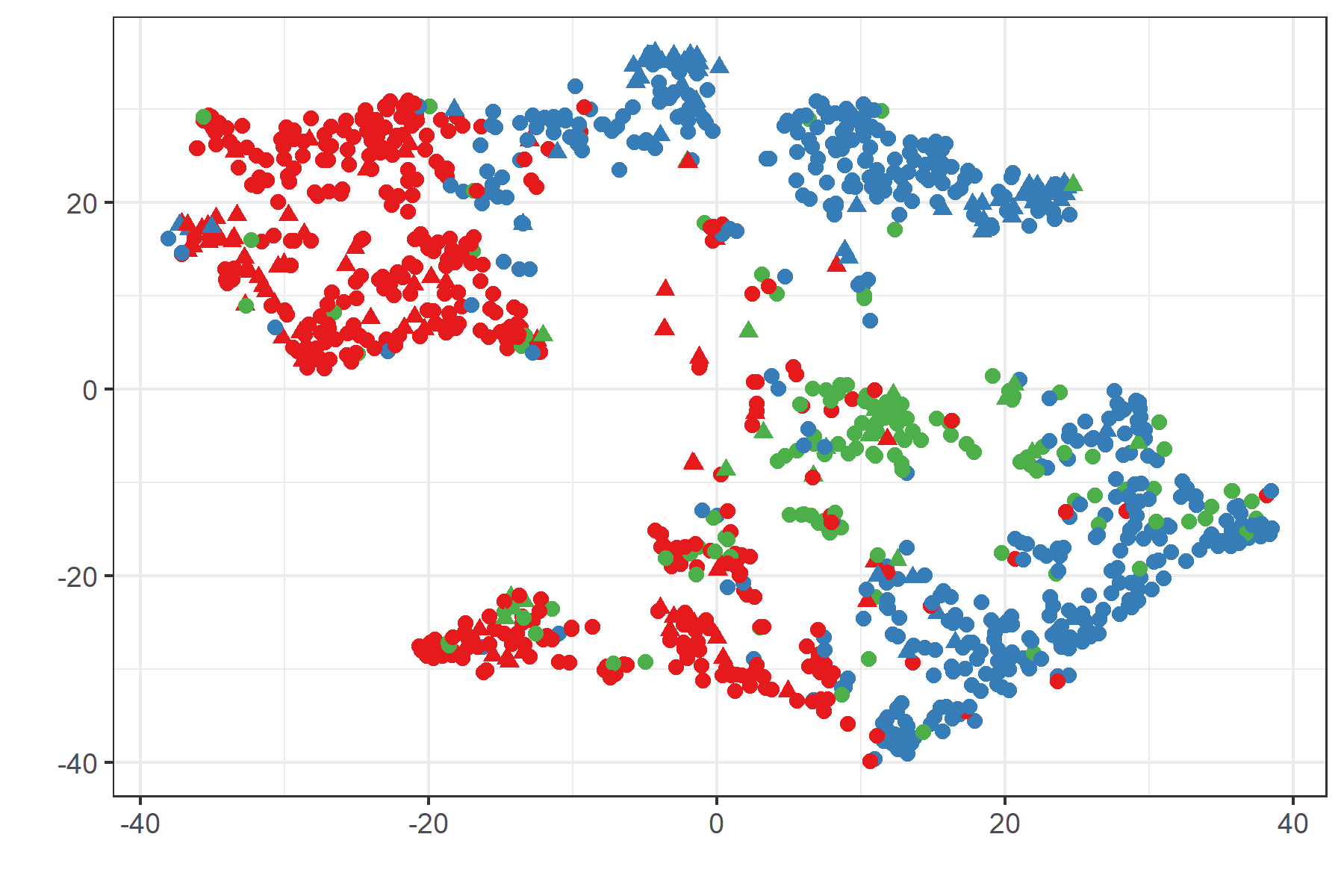}
  \caption{ LDA}
  \label{fig:bridge_sub1}
\end{subfigure}%
\begin{subfigure}{.23\linewidth}
  \centering
  \includegraphics[width=\linewidth, trim = {0 0cm 0 0cm}, clip]{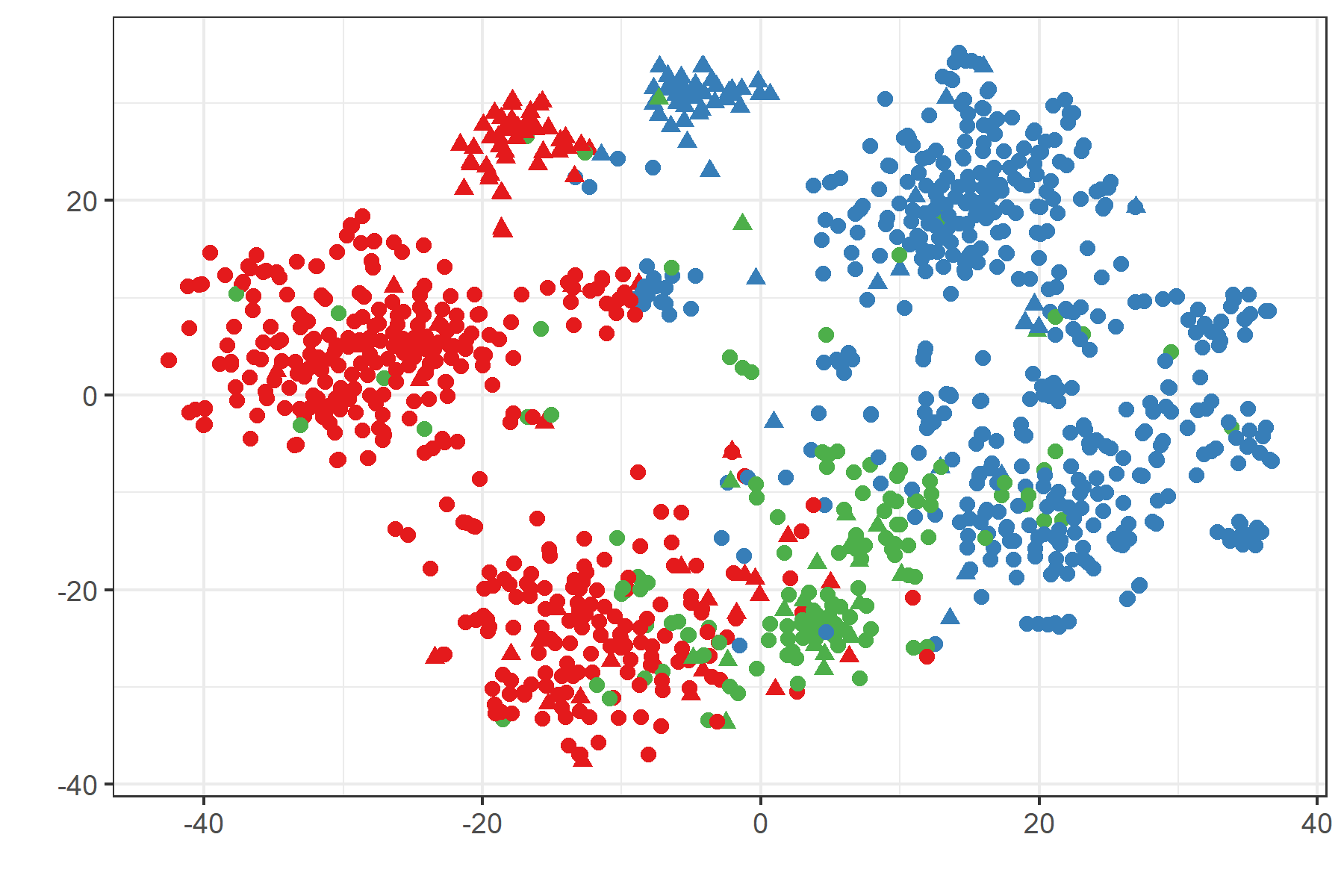}
  \caption{ node2vec}
  \label{fig:bridge_sub2}
\end{subfigure}
\begin{subfigure}{.23\linewidth}
  \centering
  \includegraphics[width=\linewidth, trim = {0 0cm 0 0cm}, clip]{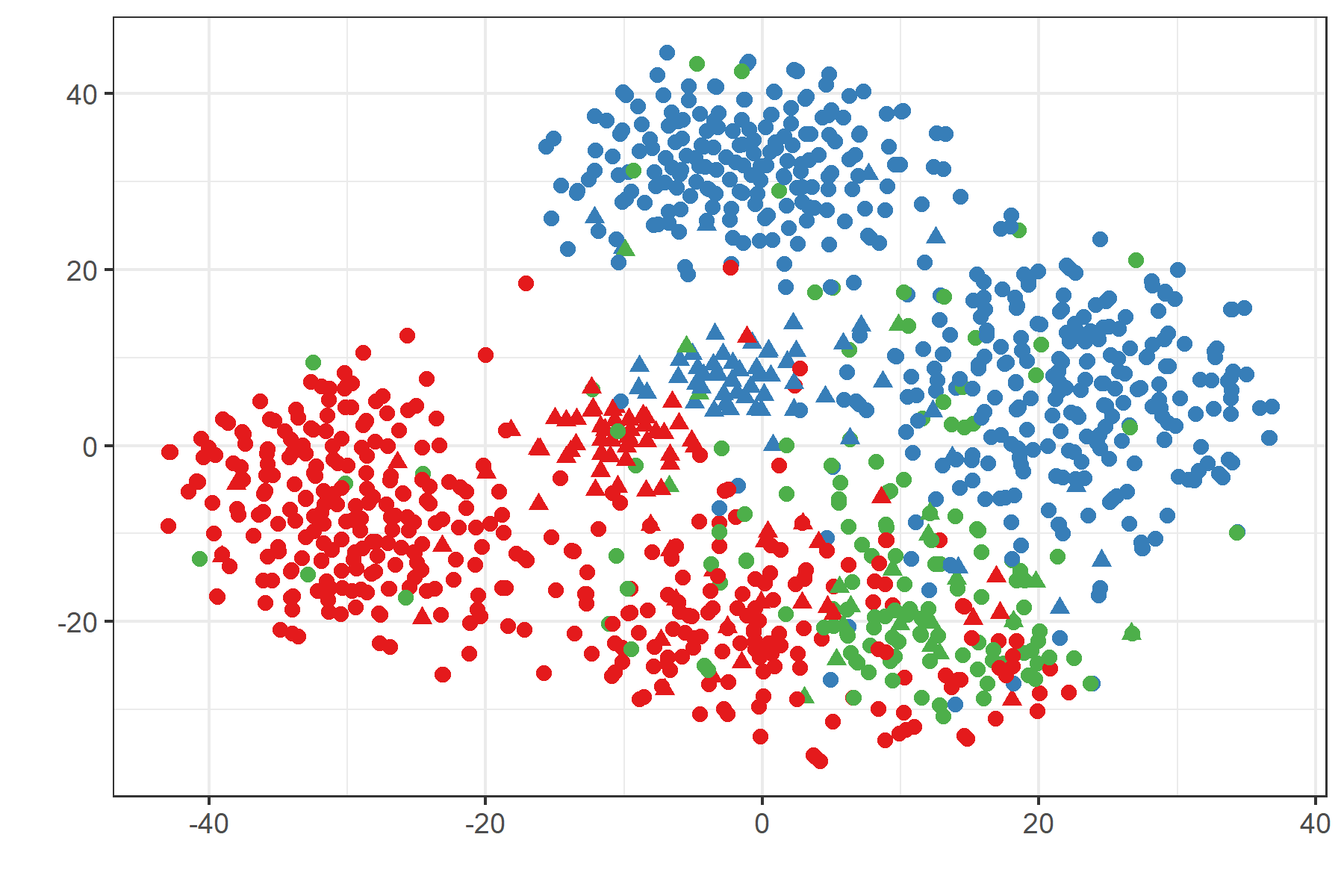}
  \caption{ BigGraph}
  \label{fig:bridge_sub3}
\end{subfigure}
\begin{subfigure}{.23\linewidth}
  \centering
  \includegraphics[width=\linewidth, trim = {0 0cm 0 0cm}, clip]{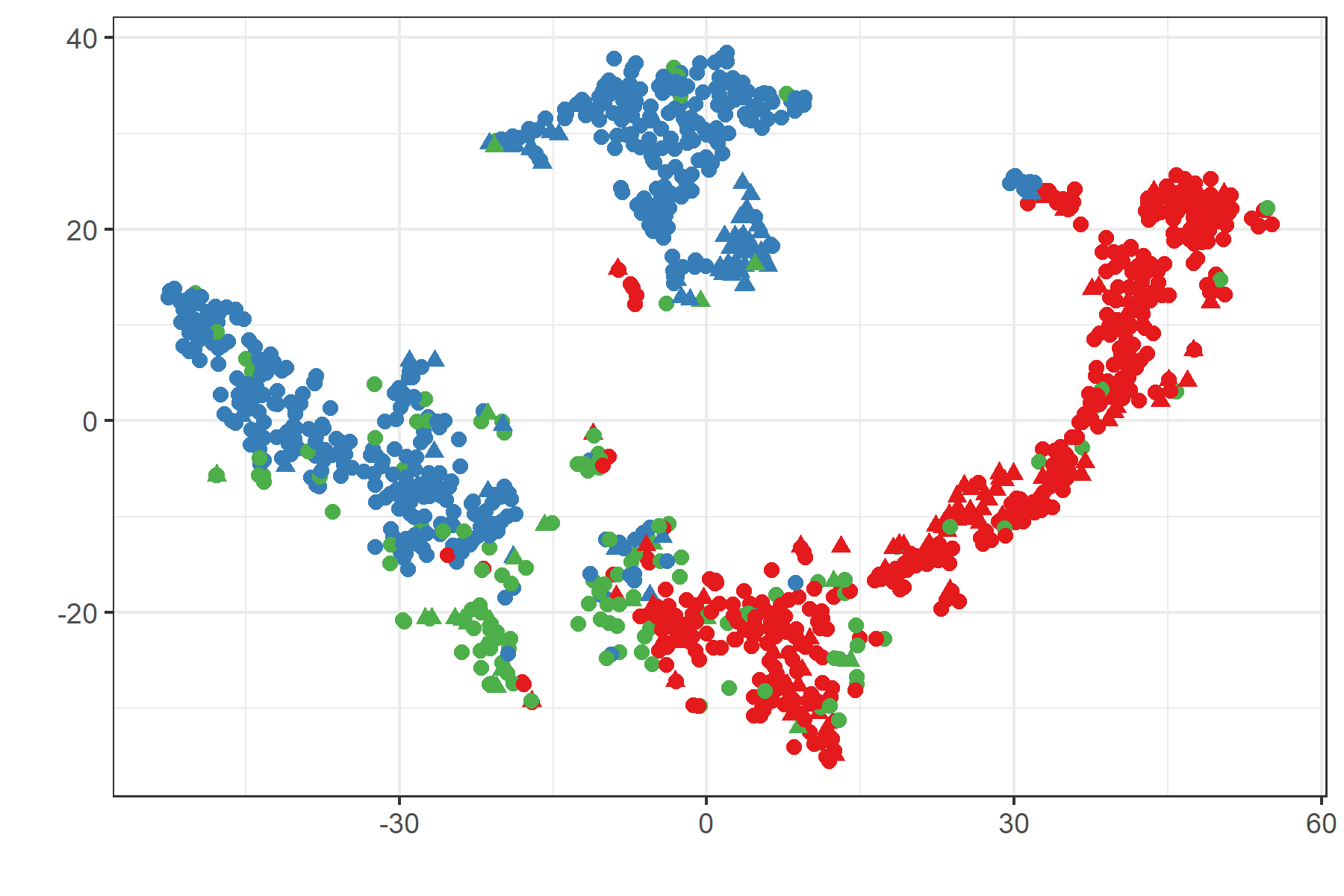}
  \caption{ GCN (w/ Features)}
  \label{fig:bridge_sub4}
\end{subfigure}
\caption{t-Distributed Stochastic Neighbor Embedding (TSNE) 2-D Visualization of US Congressional Members and Congressional Candidates for the 2018 US Midterm Elections.  Red indicates Republican politicians, Blue indicates Democrat, and Green indicates Independent.  Circle markers indicate House politicians/candidates, while triangles indicate Senate politicians/candidates.}
\label{fig:us_pol}
\end{figure*}

The data was prepared in the same way as the Yemen and IRA data, with the notable difference that the graph was constructed with \textit{friend} links as opposed to \textit{communication} links.  Only members of Congress or Congressional candidates were retained as nodes in the graph. We then used our primary models as discussed above to create \emph{content}, \emph{network}, and \emph{network + content} embeddings.  Finally, we visualized these embeddings in two dimensions using t-Distributed Stochastic Neighbor Embedding (TSNE).  The visualization of this is provided in Figure \ref{fig:us_pol}, where red indicates Republican, blue indicates Democrat, and green indicates another party.  Circles indicate House politicians, and triangles represent Senate politicians.  

From this visual validation, we see that all four models are able to capture similarity between politicians of specific parties, and within parties is generally able to separate members of the Senate from members of the House.  All four embeddings are also able to identify specific factions with each of the parties.  This visual validation gives us confidence that the selected embeddings are able to capture the rich graph and semantic features and map them to euclidean space in such a way that a k-nearest neighbors search provides value in finding similar accounts.

\section{Bot-Match Model in the Social Cybersecurity Workflow}

Bot-Match is an important tool in the social cybersecurity workflow.  While social cybersecurity workflows vary between teams and specific problem sets, a typical workflow is enumerated below.  When an information campaign is initiated or expected, social cybersecurity analysts begin developing a data collection strategy in order to collect the core data associated with the information campaign or world event.  Often this collection is either through key word filters or snowball sampling \cite{goodman1961snowball} of the network.  Most large world events require iterative collection using both keyword filtering and snowball sampling.  Once the data is collected, the team begins exploratory data analysis, which often consists of understanding the temporal and spatial density of the data as well as common hashtags and influential accounts.  With exploratory analysis complete, analysts attempt to classify accounts and images.  They often run bot/cyborg/troll detection as well as propaganda detection on the accounts.  Bots are often used as force multipliers in an information campaign, and their presence can help outline the boundary of the campaign. Additionally, analysts may attempt to classify actor type (media, politician, celebrity, etc).  The classification stage can also include meme detection to extract all memes from the social media stream (memes are helpful in that they help clearly identify messaging and target audience) \cite{BESKOW2020102170}.   By the time the team finishes classifying accounts and images, they have usually whittled the stream down to the data of interest, and now they begin a more intensive characterization of the accounts in this smaller data set.  Account characterization may be followed by campaign characterization, analysis of themes and narratives, and validation of campaign attribution (identification of the perpetrator). These generic social cybersecurity steps are summarized and enumerated below: 

\begin{enumerate}
    \item Filter social media (key word filter or snowball sampling)
    \item Exploratory data analysis (temporal/spatial distribution, common hashtags, influential accounts)
    \item Classify accounts, images, etc
    \item Characterize accounts, images, etc
    \item Characterize campaign
    \item Identify themes/messages/motives
    \item Identify target audience
    \item Attribution (identify perpetrator)
\end{enumerate}

By the time that exploratory data analysis, account/image classification and account/image characterization are complete, the team has usually found a list of sophisticated accounts that are a core part of the information campaign.  These core/interesting accounts become the input for Bot-Match, allowing the team to find similar accounts in the campaign, building out the information campaign in an iterative fashion.

In this way the Bot-Match tool and methodology is designed to be used in tandem with supervised machine learning models.  Supervised models such as Botometer \cite{davis2016botornot} and Bot-Hunter \cite{beskow2018bot} are able to find large concentrations of bots, triage the network, and provide macro level bot statistics.  However, many bots, often the most interesting and effective bots, remain undetected. This is caused by the fact that supervised models can be brittle and are biased by their training data toward specific bot types and genres \cite{beskowOSINT2020,yang2019scalable}.  These accounts are often found through exploratory data analysis.  Once found, Bot-Match allows the analyst to find other similar accounts that have also likely avoided detection.  The interesting accounts that Bot-Match returns to the analyst become new seed nodes, resulting in a recursive search pattern that allows an analyst to rapidly uncover sophisticated information operations in a matter of hours.  This method of query is more effective than the key-word boolean search that is traditionally offered in social analytics tools.  A query with all information (content and connections) is more useful than a query with a single relevant hashtag.  

The embedding type (\emph{content}, \emph{network}, and \emph{network + content}) is primarily selected based on user requirements.  If an analyst wants to find accounts that post similar content as a seed account, regardless of where they are in the network, then they should leverage semantic similarity. If trying to find nodes that are proximate in the network structure, then network embedding is more appropriate.  As a default, we found that embedding  \emph{network + content} with GCN (with Features) is a good default model if computationally feasible.  

The most attractive attribute of Bot-Match is its ability to adapt to any problem or search requirement without labeling and training a new supervised machine learning model.  All that is needed is a seed node and a target data set to search in.  This provides tangible value to social cybersecurity analysts in particular and social media analysis in general. 

In many ways the Bot-Match methodology provides a recommender system for social cybersecurity.  Item-item recommendation systems (also known as collaborative filtering) recommend items based on similarity between the items, often measured by user ratings of those items.  If you are interested in a hammer, then the recommendation system may recommend a hand saw based on item similarity.  In our case, Bot-Match says that if you are interested in a certain account manipulating a target subculture, then you may also be interested in these additional $k$ accounts that have similar connections and narratives. Selection of seed accounts could be done explicitly, or may be assumed through browsing and other search and exploratory actions.

\section{Case Study}

In this section we will illustrate the use of the Bot-Match algorithm and methodology in a social cybersecurity case study.  In this case study, we will focus on analyzing information operations and specific suspicious accounts in the 2019 Canadian National elections.   The 2019 Canadian National Elections were held on 21 October 2019.  The formal campaign started on 11 September 2019, with a total campaign duration of 40 days.  

Given the documented foreign influence in the 2016 US Elections \cite{mueller2019report,diresta2018tactics}, the Canadian authorities took extra precautions to prevent similar tampering in their national election.  The biggest policy they implemented was requiring all companies that have political advertising to set up a public facing registry with the specific ad and the name of the person who authorized that ad.  Many companies (Reddit, Google, Microsoft, others) decided to ban political advertising altogether, while others (Facebook, Instagram, CBC.ca) began setting up registries \cite{MostofCa86:online}.  This policy, while helpful in stopping manipulative paid content that the IRA leveraged in the 2016 US Election, does not stop manipulation by accounts that produce content that is not promoted through advertisement funding.  This meant that bot, troll, cyborg, and sock-puppet accounts were still able to manipulate the conversation.  Our goal was to find and analyze these malicious accounts.  

\begin{figure}[htb]
\centering
  \includegraphics[width=0.7\columnwidth, ]{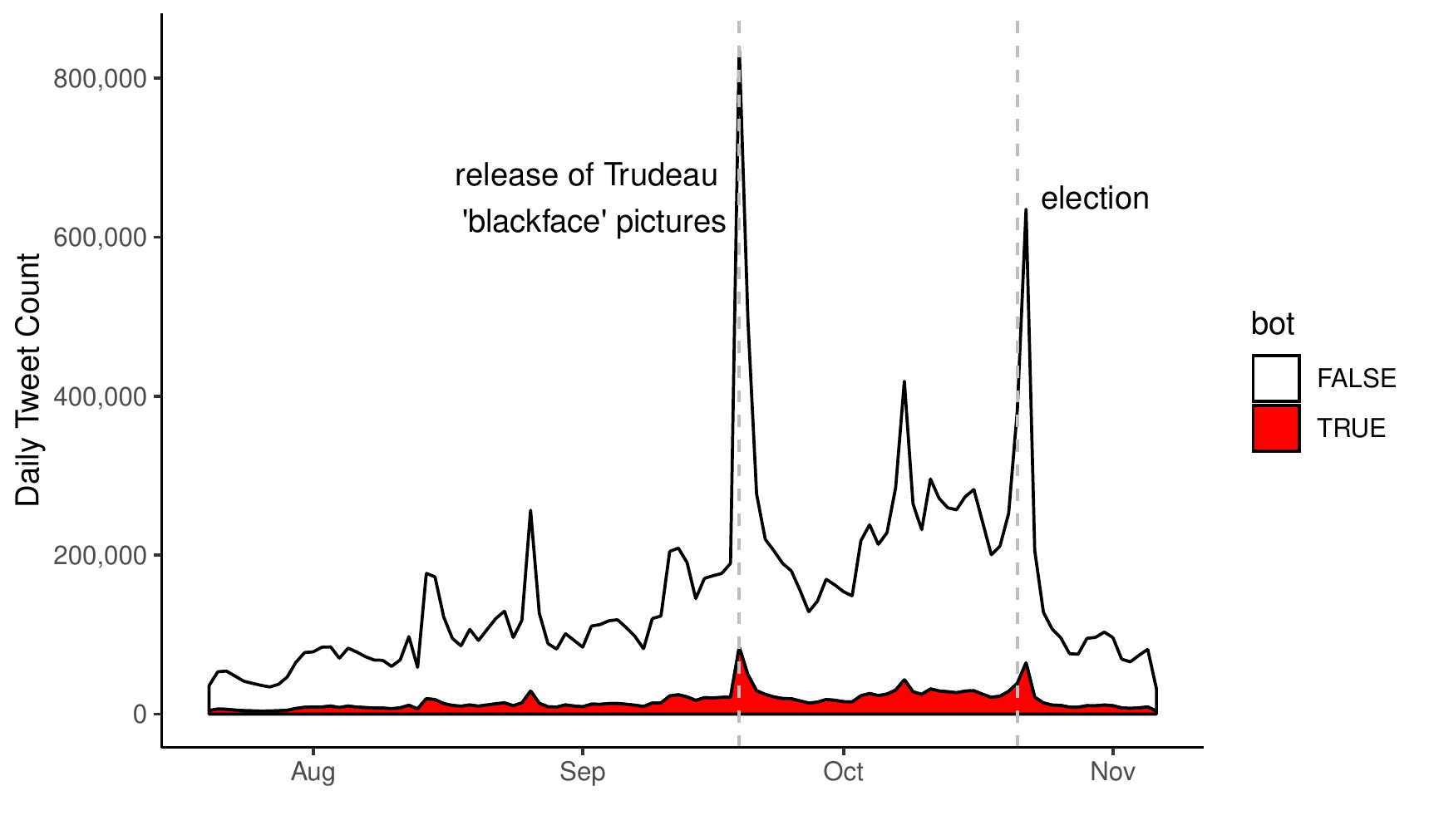}
  \caption{Temporal distribution of Canada 2019 National Election related tweets that were collected with the Twitter Streaming API.  The density of accounts with ``bot-like'' attributes as predicted by the Bot-hunter tool \cite{beskow2018bot} is shown in red.}
  \label{fig:timeline}
\end{figure}

To collect data associated with the 2019 Canadian National Election, our team used Twitter's streaming API to filter all tweets that contained hashtags associated with this event.  We did this by starting with a few general hashtags associated with the election (i.e. \#elxn43, \#cdnpoli), and then weekly adding additional trending hashtags that we found in the data, finishing with 27 hashtags or tokens associated with the election.  This collection produced 16.7 million tweets from 1.3 million unique accounts.  The temporal distribution of the data with several major news events is provided in Figure \ref{fig:timeline}.

Having collected the data, we built two embeddings for the data, one focused on the node embedding of the graph, and the other focused on content embedding for the content of the tweets.  The scale of this data collection meant that some of the embedding techniques we explored were computationally difficult or impossible as implemented above.  Given this, we used the Pytorch Biggraph model to embed the graph, and Latent Dirichlet Allocation (LDA) model to embed the content.  

The Pytorch Biggraph model was used to embed the communication network created by directed links associated with the communication modes in Twitter (mention, retweet, reply).  Latent Dirichlet Allocation (LDA) was used to embed the content.  We found the Biggraph model was more computationally tractable (20 minutes vs. 2 days for LDA), but LDA provided more meaningful nearest neighbor relationships (the Biggraph embeddings provided too much noise in returned nearest neighbors). 

Using the LDA model, we used Bot-Match methodology to find the 10 nearest neighbors of two sophisticated bot accounts that were manipulating Canadian political discussion on Twitter.  One of the bot accounts was manipulating the political right (Canadian Conservative Party) and the other was manipulating the political left (Canadian Liberal Party).  While this paper doesn't provide identifying information of the accounts, general descriptive information for both queries is illustrated and provided in Table \ref{tab:query}.  This table includes general information associated with the accounts (number or tweets, number of followers, etc), as well as bot prediction probabilities by two production supervised detection algorithms: Bot-hunter \cite{beskow2018bot} and Botometer \cite{davis2016botornot}.  It also includes top hashtags by the query accounts and nearest neighbors to evaluate semantic correlation.  

From Table \ref{tab:query} we see that all nearest neighbors of both queries are clearly associated with the stance and narrative of the query account as noted in the top hashtags.  We also see that many of the accounts appear to have some automated activity (are a bot or cyborg account) as indicated by high volume and high retweet percentages.  We also notice that many of these accounts were not detected by the state of the art production bot detection algorithms. The discrepancies between these two models, seen particularly in the second query, is likely due to very different bot genres used for training data.   

As the analyst explores these accounts, additional accounts of interest may surface, creating new Bot-Match queries, which results in the recursive nearest neighbors search of accounts of interest. This recursive nearest neighbor search of graph and semantic embedding provides an important tool for social cybersecurity practitioners.

\begin{table}[htb]
\centering
\caption{Descriptive Results of Bot-Match Queries of Sophisticated Bots Manipulating 2019 Canadian Political Parties.}
\resizebox{\linewidth}{!}{%
\begin{tabular}{lccccccl}
\toprule
\multicolumn{8}{c}{Nearest Neighbors Query with Sophisticated Conservative Bot} \\
\midrule
     Screen Name &  \# Tweets &  \# Followers &  \# Friends &  Bot-hunter &  Botometer & Retweet \% &                                       Top Hashtags \\
\midrule

       \textbf{Query} &       \textbf{39,396} &      \textbf{11,157} &     \textbf{4,616} &      \textbf{0.690} &    \textbf{ 0.148} &              \textbf{0.636} &     \textbf{TrudeauMustGo, cdnpoli, elxn43, DefundCBC, CPC} \\ 
 Neighbor 1 &       41,212 &       3,635 &     4,992 &      0.457 &     0.691 &              0.678 &  TrudeauMustGo, Trudeau, Canada, cdnpoli, Canadians \\
   Neighbor 2 &        7,886 &        935 &     1,715 &      0.378 &     0.148 &              0.592 &  TrudeauMustGo, cdnpoli, elxn43, LiberalsMustGo, SayNoToGlobalism \\
       Neighbor 3 &        7,828 &        155 &      468 &      0.280 &     0.103 &              0.737 &  TrudeauMustGo, cdnpoli, elxn43, blackface, BREAKING \\
      Neighbor 4 &       12,632 &        460 &      423 &      0.370 &     0.071 &              0.633 &  elxn43, TrudeauMustGo, cdnpoli, TrudeauWorstPM, TrudeauBlackface \\
 Neighbor 5 &        5,939 &        385 &      256 &      0.398 &     0.129 &              0.546 &  TrudeauMustGo, cdnpoli, elxn43, LiberalsMustGo, ButtsMustGo \\
 Neighbor 6 &         686 &         33 &      117 &      0.135 &     0.103 &              0.580 &  TrudeauMustGo, elxn43, cdnpoli, brownface, NotAsAdvertised \\
    Neighbor 7 &        9,562 &        406 &      858 &      0.479 &     0.083 &              0.611 &  TrudeauMustGo, elxn43, cdnpoli, elxn2019, Scheer4PM \\
      Neighbor 8 &       22,057 &        319 &      538 &      0.339 &     0.096 &              0.605 &  cdnpoli, TrudeauMustGo, LiberalsMustGo, elxn43, FakeNewsMedia \\
  Neighbor 9 &          11 &         25 &        8 &      0.477 &     0.969 &              0.705 &  cdnpoli, elxn43, TrudeauMustGo, chooseforward \\
  \midrule
  \multicolumn{8}{c}{Nearest Neighbors Query with Sophisticated Liberal Bot} \\
\midrule
     Screen Name &  \# Tweets &  \# Followers &  \# Friends &  Bot-hunter &  Botometer & Retweet \% &                                       Top Hashtags \\
\midrule
         \textbf{Query} &      \textbf{264,783} &      \textbf{16,503} &    \textbf{18,098} &      \textbf{0.470} &     \textbf{0.355} &              \textbf{0.753} &     \textbf{cdnpoli, elxn43, ChooseForward, topoli, onpoli} \\
 Neighbor 1 &       11,791 &        394 &      160 &      0.335 &     0.071 &              0.816 &  cdnpoli, elxn43, ChooseForward, NeverScheer, TeamTrudeau \\
        Neighbor 2 &       96,401 &        624 &     1,615 &      0.605 &     0.111 &              0.809 &         cdnpoli, elxn43, BREAKING, Scheer, Trudeau \\
         Neighbor 3 &       10,432 &        206 &     1,528 &      0.740 &     0.083 &              0.842 &    cdnpoli, elxn43, ChooseForward, Elxn43, CDNpoli \\
      Neighbor 4 &       35,557 &        735 &      379 &      0.560 &     0.071 &              0.870 &       cdnpoli, elxn43, ChooseForward, Trudeau, CPC \\
      Neighbor 5 &       22,937 &        226 &      983 &      0.668 &     0.138 &              0.848 &  cdnpoli, elxn43, ChooseForward, ChooseForwardWithTrudeau, IStandWithTrudeau \\
     Neighbor 6 &       27,377 &       1041 &      557 &      0.592 &     0.096 &              0.878 &  cdnpoli, elxn43, ChooseForward, TeamTrudeau, IStandWithTrudeau \\
          Neighbor 7 &        5,026 &        438 &     1,540 &      0.558 &     0.103 &              0.843 & cdnpoli, elxn43, ChooseForward, ScheerWasSoPoorThat, IStandWithTrudeau \\
      Neighbor 8 &        3,445 &        334 &     1,026 &      0.564 &     0.066 &              0.768 &  cdnpoli, elxn43, ChooseForward, NeverScheer, YankeeDoodleAndy \\
           Neighbor 9 &       37,845 &        257 &      387 &      0.435 &     0.222 &              0.927 &        cdnpoli, elxn43, ChooseForward, CPC, onpoli \\
\bottomrule
\end{tabular}}
\label{tab:query}
\end{table}

As discussed above, we have already deployed prototype models of Bot-Match to monitor malicious disinformation in the Canadian elections.  Having discovered sophisticated bot accounts manipulating both the political left and political right in Canada, we used Bot-Match to build out these campaigns and delineate the respective manipulative (dis)information operations.

\section{Uses Beyond Social Cybersecurity}

The concept of using an account and all associated features and connections as a query has many applications beyond social cybersecurity.  These applications include retail, link prediction, intelligence, and information retrieval.  

The retail business is one of the first adopters of recommender systems, and is arguably the most mature at deploying scalable collaborative filtering.  These systems are inherently constrained by the user, and suffer from cold-start challenges.  All product recommendations for the user are limited by the users own biases and ignorance, some of which they'd like to circumvent.  Using the concept of an account query, an online retailer could allow a user to receive recommendations based on someone else's account (a celebrity, a friend, or someone else whose tastes they admire and wish to emulate).  Social recommendations are some of the most powerful product recommendations, and allowing a person to receive recommendations as if they were someone else could be profitable for the retail industry.  This approach would have a number of privacy hurdles to overcome, but if implemented correctly, allowing a customer to ``Shop as if they were ....'' could be the next big step in online retail.

Social media companies use recommendation systems and link prediction algorithms to make recommendations to a user based on their interests, content, and existing links.  As with the retail business, some users may like to see recommendations as if they were someone else.  For example, a young professional would like to see recommendations being presented to an established professional in their field who they follow and attempt to emulate.  Once again, these raise significant but not insurmountable privacy concerns.  

The final application area is in the area of intelligence.  Many systems used for the intelligence community use simple boolean search patterns to search repositories of unstructured data.  While key word searches may be required for initial exploration, as analysts find entities they are interested in, these entities and all content and connections associated with them could be used as a search query.  This type of rich query could provide better search results for intelligence analysts.

\section{Conclusions}

This paper evaluates state of the art graphical and semantic embedding for social media data, and then leverages these embeddings for bot detection to enable social cybersecurity.  Bot-Match is evaluated in two new social cybersecurity datasets, validated on a third dataset associated with US politics, and then demonstrated on a fourth dataset associated with the 2019 Canadian National Elections.  Within the emerging discipline of social cybersecurity, the Bot-Match paradigm provides a novel way for analysts to find similar nefarious actors and recursively discover a complex disinformation operation without labeling and training a supervised machine learning model.  As such, it extends the concept of content based information retrieval beyond multi-media.  Finally, while used within the social cybersecurity context, this approach has broad application to retrieval tasks that are characterized by network connections and semantic content. This includes document retrieval, recommendation systems, social recommendation, and other use cases.

\begin{acks}
This work was supported in part by the Office of Naval Research (ONR) Multidisciplinary University Research Initiative Award  N000140811186, Award N000141812108, ONR Award N00014182106 and the Center for Computational Analysis of Social and Organization Systems (CASOS). The views and conclusions contained in this document are those of the authors and should not be interpreted as representing the official policies, either expressed or implied, of the ONR or the U.S. Government.

\end{acks}

\bibliographystyle{ACM-Reference-Format}
\bibliography{references.bib}










\end{document}